\documentclass[pre,twocolumn,floatfix,showpacs,showkeys]{revtex4}
\usepackage{hyperref}
\usepackage{amssymb}
\usepackage{amsfonts}
\usepackage{graphicx}
\usepackage{aecompl}
\usepackage{ae}

\newcommand{\Eq}[1]{Eq.~(\ref{eq:#1})}
\newcommand{\Fig}[1]{Fig.~\ref{fig:#1}}

\begin{document}

\title{
  Piling and avalanches of magnetized particles
}

\author{
  S. Fazekas$^{1,2}$, J. Kert\'esz$^{1}$, and D. E. Wolf$^{3}$
}

\affiliation{
  $^1$Department of Theoretical Physics,\\
  Budapest University of Technology and Economics,\\
  H-1111 Budapest, Hungary\\
  $^2$Theoretical Solid State Research Group
  of the Hungarian Academy of Sciences,\\
  Budapest University of Technology and Economics,\\
  H-1111 Budapest, Hungary\\
  $^3$Institute of Physics, University Duisburg-Essen,\\
  47048 Duisburg, Germany
}

\date{April 7, 2004}

\begin{abstract}
  We performed computer simulations based on a two-dimensional 
  Distinct Element Method to study granular systems of magnetized 
  spherical particles. We measured the angle of repose and the 
  surface roughness of particle piles, and we studied the effect 
  of magnetization on avalanching. We report linear dependence 
  of both angle of repose and surface roughness on the ratio $f$ of 
  the magnetic dipole interaction and the gravitational force
  (\emph{interparticle force ratio}).  There is a  
  difference in avalanche formation at small and at large 
  interparticle force ratios. The transition is at $f_c \approx 7$. 
  For $f < f_c$ the particles forming the 
  avalanches leave the system in a quasi-continuous granular flow 
  (\emph{granular regime}), while for $f > f_c$ the avalanches are 
  formed by long particle clusters (\emph{correlated regime}).
  The transition is not sharp. We give plausible estimates for $f_c$
  based on stability criteria.
\end{abstract}




\maketitle

\bigskip 

\section{Introduction} 

The dipole interaction between magnetized particles can be viewed
as an anisotropic adhesion force. Because its strength can be easily 
manipulated by the strength of the magnetizing field, magnetized 
particles have been recently proposed 
\cite{forsyth-pre01,hutton-phd02}
to get insight into the 
transition from noncohesive to cohesive grains.
Previously adhesion effects have mainly been studied
in form of moisture-induced changes,
significant for industrial processes in fields as 
pharmaceuticals, agriculture, and constructions.

Some time ago Hornbaker et al. \cite{hornbaker-nat97} 
addressed the question
how sand castles stand. They stated that already 
small quantities of wetting liquid can dramatically
change the properties of granular media, leading to 
large increase in the angle of repose and 
correlation in grain motion.
Theoretical studies on the angle of repose 
based on stability criteria have been done 
by Albert et al. \cite{albert-pre97}. 
They theoretically determined the dependence of the angle of repose 
on cohesive forces, and applied the results to wet granular material.

Experimental studies of Tegzes et al. on angle of repose 
using the draining crater method \cite{tegzes-pre99} 
and on avalanches using a rotating drum apparatus 
\cite{tegzes-pre03} identify 
three distinct regimes as the liquid content is increased:
a \emph{granular regime} in which the grains move individually, 
a \emph{correlated regime} in which the grains move in correlated 
clusters, and a \emph{plastic regime} in which
the grains flow coherently.

Experiments of Quintanilla et al. \cite{quintanilla-prl01}
using the rotating drum apparatus address the question of 
self-organized critical behavior in avalanches of slightly 
cohesive powders. Their results show that avalanche sizes 
do not follow a power-law distribution, however, they scale with 
powder cohesiveness.
Samadani et al. \cite{samadani-pre01} 
studied the effect of interstitial fluid 
on the angle of repose and the segregation   
of granular matter poured into a quasi-two-dimensional silo.

To study the transition from noncohesive to cohesive behavior,
Forsyth et al. \cite{forsyth-pre01,hutton-phd02},
adopting the widely suggested idea that competition between the 
interparticle forces and the inertial forces determines the
behavior of cohesive granular materials,
suggested a method based on magnetized particles.
The particles placed 
in an external magnetic field become magnetized, 
all having the same magnetic orientation 
parallel to the field.
Varying the strength of the field allows to continuously vary 
the resulting interparticle magnetic force.
Using nonmagnetic perspex walls the particle-wall interaction
remains unchanged relative to the noncohesive state.
Using particles under same packing conditions 
it is ensured that the initial conditions are as 
uniform as possible.

We carried out computer simulations on a system corresponding 
to the experiments of 
Forsyth et al. \cite{forsyth-pre01,hutton-phd02}
and studied the angle of repose, the surface roughness,
and the effect of magnetization on avalanching in two-dimensional
particle piles.

The magnetic interaction of magnetized grains is highly anisotropic,
and the fixed external field introduces even more anisotropy
as the grains are aligned to the field. 
A similar experimental setup \cite{szalmas-bme00}, 
but with particles carrying a 
remanent magnetization in the absence of an external magnetic 
field, would partly diminish the mentioned anisotropy, however
in this case the magnetizations and the interparticle forces
are not as well defined as in the experiments of 
Forsyth et al. \cite{forsyth-pre01,hutton-phd02}. 

Because of the strong anisotropy and the longer interaction range one
can expect differences between the results on magnetized particles and
wet granular systems, however the basic effects should be the same.

\section{Simulation method} 

We performed computer simulations based on a
two-dimensional Distinct Element Method (DEM) 
\cite{cundall-geo79} (for a review see
\cite{ristow-arcph94,herrmann-cmt98,luding-book04}
and references therein)
to study granular systems of magnetized spherical particles. 
The particles are magnetized by a
constant external field, all having the same magnetic orientation 
parallel to the field. 
The magnetization is modeled with dipoles.
We neglect any coupling between the
magnetic orientation and particle rotation (i.e., the particles
can rotate freely, while their magnetic dipole is fixed).

For characterizing the strength of the interparticle force, we 
introduce a dimensionless quantity defined by the ratio of 
the maximum magnetic 
interparticle force at contact and the gravitational force.

The magnetic force acting on a dipole ${\bf m}_{2}$ 
situated at distance $r_{21}$ from a dipole ${\bf m}_{1}$,
along the direction ${\bf n}_{21}$ is given by

\begin{eqnarray}
{\bf F}_{21}
  = 
\frac{\mu_{0}}{4\pi}\ \frac{3}{r_{21}^{4}}\ 
\left[\ 
      ( {\bf n}_{21} {\bf m}_{2} )\ {\bf m}_{1}
      +\ ( {\bf n}_{21} {\bf m}_{1} )\ {\bf m}_{2}
\right.           
\nonumber \\
\left.
      - 5\ 
        ( {\bf n}_{21} {\bf m}_{1} )\ 
        ( {\bf n}_{21} {\bf m}_{2} )\ 
          {\bf n}_{21}
      + ( {\bf m}_{1} {\bf m}_{2} )\ 
          {\bf n}_{21}
    \ \right].
\end{eqnarray}

For identical hard spherical particles of diameter $D$ 
and magnetic dipole $S$, the largest possible dipole-dipole magnetic
force is

\begin{equation}
F_m = \frac{\mu_{0}}{4\pi}\ \frac{6S^2}{D^{4}},
\end{equation}

\noindent which corresponds to a head-to-tail
configuration, the dipoles having the same orientation.

We define the magnetic \emph{interparticle force ratio} as

\begin{equation}
f = F_m / F_g, 
\end{equation}

\noindent where $F_g = mg$ is the gravitational force 
($m$ denotes the particle mass and $g$ is the gravitational 
acceleration).

Considering mass density $\rho$ and magnetization $M$, we have
$m = \rho V$, $S = MV$, $V = \pi D^3 / 6$, and thus

\begin{equation}\label{eq:f_def}
f = \frac{F_m}{F_g} 
  = \frac{\mu_{0}}{4\pi}\ \frac{6S^2}{m g D^{4}} 
  = \frac{\mu_{0}}{4\pi}\ \frac{\pi M^2}{\rho g D}.
\end{equation}

Assuming some $f$ interparticle force ratio, 
from the previous equation the corresponding magnetization 
can be calculated as

\begin{equation}
M = \left( f ~ \frac{4\pi}{\mu_0} \frac{\rho g D}{\pi} \right)^{1/2}.
\end{equation}

In our simulations we used $\rho = 7.5\ g/cm^3$ (which
corresponds approximately to the mass density of steel),
$g = 9.8\ m/s^2$, and interparticle force ratio $f < 25$.

The diameter of the spherical particles was
taken from the $0.7-0.9\ mm$ interval, with a \emph{Gauss-like}
distribution having the mean of $0.8\ mm$. 
The vast majority of particles had diameters very close to
$D = 0.8\ mm$. The \emph{Gauss-like}
distribution is given by the average of $4$ independent uniformly
distributed random variables in the mentioned interval. This slight
polydispersity, resembling real experimental setups,
is used to avoid effects 
originated from symmetries of monodisperse systems.

The long range magnetic interaction is taken in consideration
within a reasonable cutoff distance as a dipole-dipole interaction.
We choose the magnetic interaction cutoff to $6.25 D$ (where $D$ is
the average particle diameter). As shown in a previous study 
\cite{fazekas-pre03}, $5 D$ value already gives a reasonable 
magnetic interaction cutoff in two-dimensional dipolar hard 
sphere systems regarding the local ordering.
The angle of repose, the surface roughness, and the particle
avalanches depend crucially on local orderings inside the
pile, as noted for example by Altshuler et al. in 
\cite{altshuler-prl01}. 
The used cutoff keeps the character of local orderings and 
changes the magnetic energy per particle by less than $5\%$
\cite{fazekas-pre03}.

We calculate the
collision interaction of particles using the 
Hertz contact model \cite{landau-70}
with appropriate damping
\cite{kuwabara-jjap87}. 
We implement Coulomb
sliding friction for large relative translational velocities
and for numerical stability
viscous friction for small velocities, with 
continuous transition between the two, controlled by a 
(large) viscous friction coefficient. We do not use
any static or rolling friction model. A grid based method 
is used to identify neighboring (and potentially 
colliding) particle pairs. 

The parameters of the Hertz contact model were chosen such that they
correspond to Young modulus of approximately $0.015\ GPa$ and
restitution coefficient of approximately $0.86$.  These are
characteristic values for hard rubber elastomers (used for example in
constructing golf ball covers). The Young modulus is orders of
magnitudes smaller than that of steel, a choice enabling
realistic CPU times with the DEM method.  The particle-particle and
the particle-wall sliding friction coefficient was $0.5$ and $0.7$
(characteristic for steel-steel and steel-perspex friction).

The translational motion of
particles is integrated based on Newton's equation using 
Verlet's leap-frog method. The rotational state of particles
is integrated with Euler's method.
The integration time step was $5\ \mu s$.
With the used elastic parameters 
a good lower estimate for collision times is $170\ \mu s$.
In such conditions,
the used integration time step gave good numerical stability 
and also fairly good response time on
PCs with $1.8$ GHz CPUs available at the time of writing.

\begin{figure}[tbhp]
\begin{center}
\begin{tabular}{r}
\includegraphics{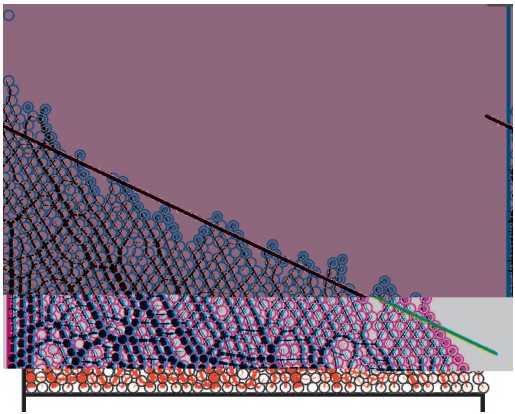} \\
\end{tabular}
\end{center}
\caption{
  Simulation setup. The particles are introduced 
  with constant rate one by one 
  at small random distances from the left wall.
  The particles can leave the system on the right side. 
  The surface angle (i.e. the angle of repose)
  is measured by fitting a straight line over the positions of
  the surface particles (marked with black). 
  The figure also shows the normal contact forces. 
  The thickness of the lines connecting the centers of 
  the particles in contact is proportional to the 
  normal contact force.
  The sample corresponds to $f = 6$ interparticle
  force ratio. For color figure see the electronic version.
}
\label{fig:syssam}
\end{figure}

The simulation setup can be seen in \Fig{syssam}. 
The external magnetic field is vertical.
The particles are added one by one with constant rate along vertical
trajectories at small (maximum one particle diameter) random distance
from the left wall. They either reach the pile with a given velocity 
(i.e. \emph{they are fired into the pile}), or their impact velocity
is set to zero (i.e. \emph{they are placed gently on the pile}).

The system's bottom wall is sticky. Any particle touching
the bottom wall sticks to the wall. This builds up a
\emph{random base} (see the experimental setup 
used by Altshuler et al. described in \cite{altshuler-prl01}).
The particles can leave the system on the right side.
The particles are removed from the simulation when their
distance from the bottom-right corner is larger
than the magnetic interaction cutoff distance.
The system's bottom is $51.25 D$ length. 
Only this system size was used,
finite size effects were not studied.

The \emph{surface particles} (marked with black in \Fig{syssam}) 
are identified with the \emph{weighted alpha shape algorithm}
\cite{edelsbrunner-techrep92,akkiraju-icg95}. 
Alpha shapes are
generalizations of the convex hull and can be used for
shape reconstruction from a dense unorganized set of data points.
The weighted alpha shapes are extensions of this kind of
shape reconstruction to a set of spheres (as in our case).
We used the implementation included in the
Computational Geometry Algorithms Library \cite{CGAL-2.4}.
The algorithm's \emph{alpha} parameter was set
to the square of the mean particle size.
This gave satisfactory results.
The surface angle (i.e. the angle of repose)
is measured by fitting a straight line over the positions of
the surface particles. The surface roughness is given
by the standard deviation of the surface points from the 
fitted line.

As part of our investigations, with a special \emph{side wall model}, 
we also simulated the effect of 
front and back walls in a Hele-Shaw cell geometry encountered in 
experimental studies. We took into consideration the frictional 
interaction with side walls by summing the magnitude of normal 
forces acting on one particle, directing a well defined percentage
of this \emph{pressure} on the walls, and deriving a frictional 
force using the already mentioned friction model. The percentage
of the total force directed on the side walls was a parameter of our
simulations.

We performed three sets of simulations: (a) the particles
were fired into the pile,
(b) the particles were placed gently on the pile,
and (c) the particles were fired into the pile, 
while $4\%$ of the internal \emph{pressure} was directed
on the \emph{front and back walls} (see side wall model).
In both (a) and (c) the particles reached the pile 
with $0.5\ m/s$ impact velocity, which corresponds to
approximately $16 D$ dropping height.
In all three simulation sets we executed runs at different
interparticle force ratios. In each run we started with an
empty system, and introducing $12000$
particles, one particle every $3000$ integration steps,
we numerically integrated the system for $3$ minutes (simulated
time). 

In the first part of the process the number of particles
in the system increased monotonically. 
After a pile was built, avalanches
started, which in a pulsating manner moved particles out of the 
system. In this
way the number of particles began to oscillate around some
well defined value. In this latter part we identified the 
surface particles every $500$ integration steps, 
and we measured the slope of the fitted
surface line and the standard deviation of surface points from
this line. The average of this quantities over the simulated
time gave the measured angle of repose and surface roughness.

We also measured the avalanche sizes and avalanche durations.
This can be done in many different ways.
We define an avalanche by a number of individual
events on the time scale of the integration steps
(i.e. the smallest simulation time step)
in which (at least) one 
particle leaves the system, and the time between two consecutive
events is smaller than a well defined value. We take this value
equal to the time corresponding to $3000$ integration time steps.
Our choice is based on the system's observed dynamic 
time scale, and the fact that
one new particle is introduced (i.e. the system is 
perturbed) every $3000$ integration time steps, and thus
on a larger time scale there are surely uncorrelated 
events.

\section{Simulation results} 

\subsection{Angle of repose and surface roughness} 

\begin{figure}[tbhp]
\begin{center}
\begin{tabular}{r}
\includegraphics{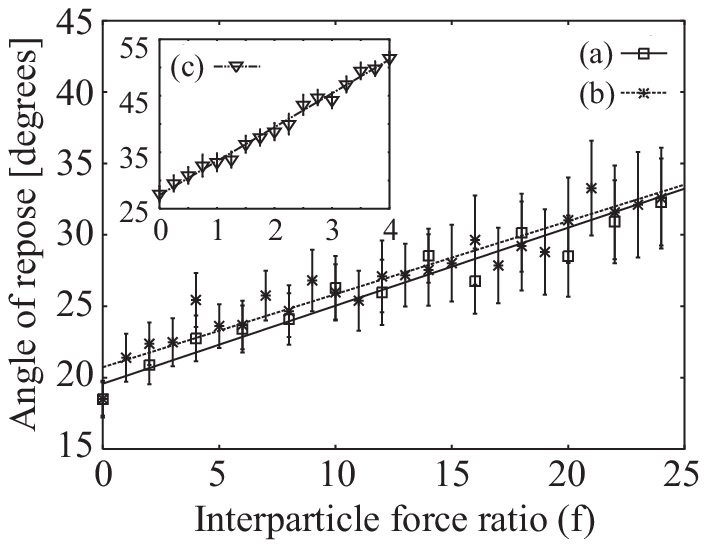} \\
\\
\includegraphics{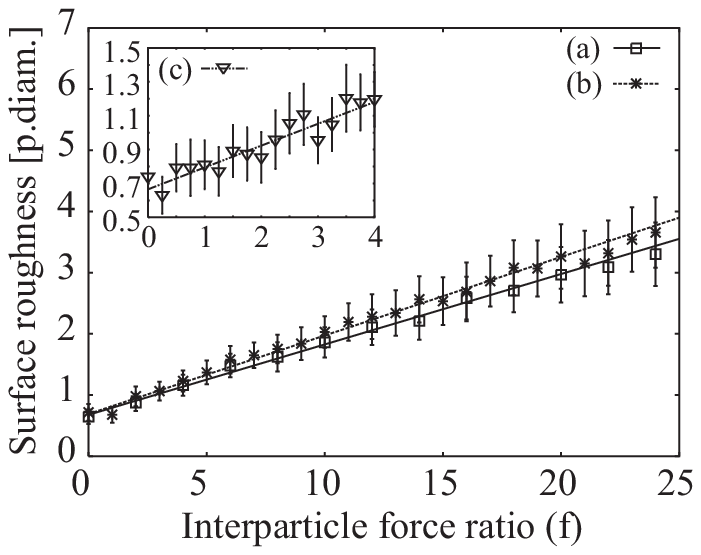} \\
\end{tabular}
\end{center}
\caption{
  Angle of repose (upper panel) and surface roughness (lower panel)
  at different magnetic interparticle 
  force ratios. The angle of repose is 
  measured in degrees. 
  The surface roughness is measured in (average) particle diameters.
  We executed three set of simulations.
  In (a) and (c) the particles were fired into the
  pile, while in (b) the particles were placed gently on the pile.
  In (c) an artificial side wall effect was switched on (see 
  text for details). 
}
\label{fig:angofrepnroughn}
\end{figure}

In all cases both the angle of repose and 
the surface roughness (in the examined domain)
exhibit a linear dependence on the interparticle force ratio
(see \Fig{angofrepnroughn}).

The angle of repose in case (a) and (b) increases by approximately 
$0.5$ degree per unit change of interparticle force ratio 
(see upper panel in \Fig{angofrepnroughn}).
This is in good accordance with the experimental results
of Forsyth et al. \cite{forsyth-pre01,hutton-phd02},
however the angle of repose at zero magnetization in
our case is about $10$ degrees smaller. This can be the 
result of the missing side wall effect 
(see for example \cite{dury-pre98}), and the missing 
static and rolling friction (see for example 
\cite{zhou-physa99,zhou-powtech02}).

At zero magnetization the average surface roughness is
about $0.7$ particle diameters, and in all cases increases 
by approximately $0.12$ particle diameters
per unit change of interparticle force ratio (see lower
panel in \Fig{angofrepnroughn}).

\begin{figure}[tbhp]
\begin{center}
\begin{tabular}{r}
\includegraphics{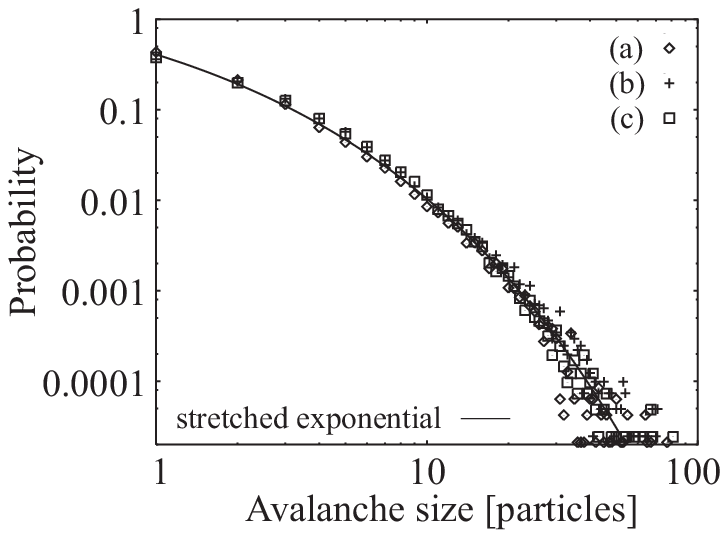} \\
\\
\includegraphics{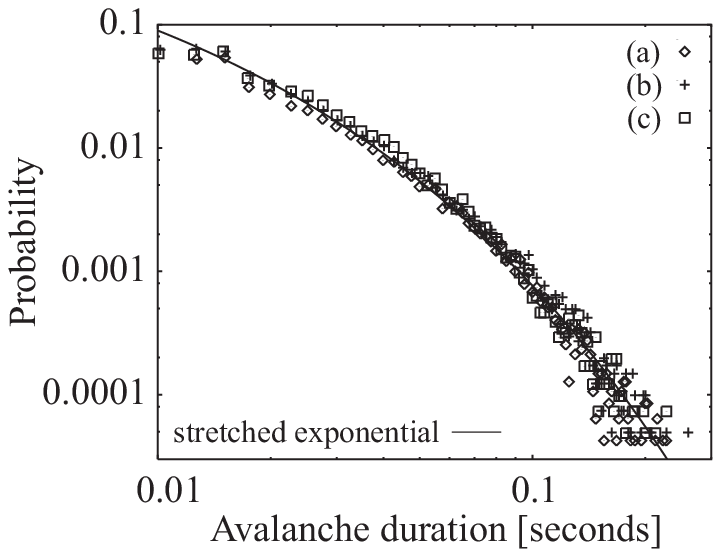} \\
\end{tabular}
\end{center}
\caption{
  Distribution of particle avalanche sizes (upper panel)
  and avalanche durations (lower panel) at zero magnetization.
  We examined three different simulation setups (see text for details).
  Firing the particles into the pile (i.e. dropping them from a given
  height) or placing them gently, and switching on or off the side
  wall effect gave no qualitative difference. Over our simulation
  data (for both avalanche sizes and avalanche durations) 
  we could fit a stretched exponential with $\gamma = 0.43$.
}
\label{fig:statlong}
\end{figure}

As a consequence of our side wall model, the angle of repose in 
case (c) is about $8$ degrees higher at zero magnetization in 
agreement with experimentally observed effects of front and back 
walls in Hele-Shaw cells. However, the way we model the side walls 
leads to a stronger increase of the angle of repose with $f$ than
in the experiments of Forsyth et al. \cite{forsyth-pre01,hutton-phd02}
(see inset of upper panel in \Fig{angofrepnroughn}).
The side wall effect does not
influence the surface roughness
(see inset of lower panel in \Fig{angofrepnroughn}).

\subsection{Particle avalanches at zero magnetization} 

We carefully examined the distribution of particle avalanches at zero
magnetization in all three simulation sets. We also executed
extra runs introducing a total of $144000$ particles,
one particle every $3000$ integration steps,
integrating for a total of $36$ minutes (simulated time). 
It must be noted that simulating at zero magnetization (i.e. without
magnetization) is about $10$ times faster, because only the
(short-range) collision interaction must be calculated.
This permitted longer simulation times.
The corresponding particle avalanche size and duration 
statistics can be seen in \Fig{statlong}.

Firing the particles into the pile (i.e. dropping them from a given
height) or placing them gently, and switching on or off the side
wall effect gave no qualitative difference. Over both the avalanche 
size and duration distribution data
we could fit stretched exponentials of form 

\begin{equation}
P(x) = P_0 \exp[-(x/x_0)^{\gamma}]
\end{equation}

\noindent with $\gamma = 0.43$.

Our $\gamma$ value is in good agreement with
the experimental results of \cite{feder-fract95}
and is the same value as the one found by Frette et al. 
\cite{frette-pre01} for piles of rice with small anisotropy.
By contrast recent work of 
Costello et al. \cite{costello-pre03} 
presents carefully collected and detailed experimental results
on piling of uniform spherical glass beads,
which show a power law behavior with an exponential cutoff.
Costello et al. argue that the exponential cutoff depends on
the height from which the particles are dropped and probably also
on cohesion forces. It was not our intention in this study to 
contribute to the clarification of this point.

\subsection{Effect of magnetization on particle avalanching} 

We analyzed the effect of magnetization
on particle avalanching, and we found that there 
is a difference in avalanche formation at small 
and at large interparticle force ratios 
(see avalanche movies \cite{fazekas-aval}). 
We identified a \emph{granular} and a 
\emph{correlated regime}. The transition 
between the two regimes is not sharp.
Similar regimes were identified experimentally by 
Tegzes et al. \cite{tegzes-pre99,tegzes-pre03}
in case of wet granular materials.
At high liquid content (i.e. large interparticle force ratio) 
they could also identify a third \emph{plastic regime}.

Studying the recordings from our simulations \cite{fazekas-aval},
it can be observed that for small magnetizations 
the avalanches are formed by small vertical chains
following each other at short times, giving the impression of
a quasi-continuous flow (\emph{granular regime}). 
As the particle magnetization increases, 
at $f_c \approx 7$ the previously quasi-continuous flow 
is replaced by individual narrow and long particle clusters
falling at the system's boundary
(\emph{correlated regime}).

\begin{figure}[tbhp]
\begin{center}
\begin{tabular}{r}
\includegraphics{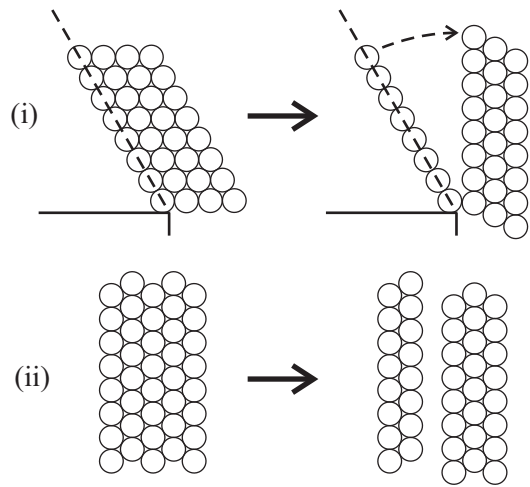} \\
\end{tabular}
\end{center}
\caption{
  Elementary processes characteristic for the dynamics of avalanche 
  formation: (i) \emph{peeling} and (ii) \emph{splitting}.
  A particle cluster composed of parallel chains can peel off the 
  pile by rotating to a more favorable orientation, and can 
  disintegrate into clusters composed of smaller number of chains.
  Combinations of these two processes determine the size of the 
  outflowing clusters.
}
\label{fig:peelnsplit}
\end{figure}

There are two elementary processes characteristic for 
the dynamics of avalanche 
formation: \emph{peeling} and \emph{splitting} (see \Fig{peelnsplit}).
Near the foot of the pile the particles are arranged approximately 
in a triangular lattice with a horizontal base. By contrast magnetic 
interactions would favor a triangular lattice with a vertical base.
Clusters of particles near the free surface of the pile 
can \emph{peel off} the pile by rotating $30$ degrees into this 
favorable configuration (see part (i) of \Fig{peelnsplit}).
A favorably oriented domain of particles on a triangular lattice
consists of vertical chains shifted by half a particle diameter from 
one chain to the next. This domain may be stable or may disintegrate 
into smaller domains or chains, depending on whether or not the
magnetic interactions are strong enough to prevent the dilation, 
which is necessary to allow for relative motion of chains within 
the domain (see part (ii) of \Fig{peelnsplit}). 

Combinations of these two processes determine the size of the 
outflowing clusters,
which typically consist of $\nu$ parallel vertical chains of length 
$N$. Both $N$ and $\nu$ depend on $f$. Taking into account only
the magnetic interaction between nearest neighbors one can give 
plausible estimates for this dependence.

We compare the gain in magnetic energy and the loss in gravitational 
energy when a cluster of particle chains of length $N$ rotates by 
an angle $\alpha$ into a position where the chains are aligned
with the magnetic field. 
The magnetic moment of each particle is
always aligned with the magnetic field, so that only the dipolar 
interaction between the grains matters. 
We take into consideration only intrachain 
interactions, neglecting the interchain interactions. 
The dipole-dipole-interaction
potential between two particles whose center of mass connection 
is tilted by an angle $\alpha$ with respect to the magnetic field
is

\begin{equation}
E_{m} = 
  \frac{\mu_{0}}{4\pi}\ \frac{S^2}{D^{3}}\ 
    \left( 1 - 3 \cos^2\alpha \right).
\end{equation}

\noindent Hence rotating them into the field direction ($\alpha = 0$)
lowers the energy per particle in the cluster by an amount 
proportional to

\begin{equation}
\Delta E_{m} = 
  \frac{\mu_{0}}{4\pi}\ \frac{3 S^2}{D^{3}}\ 
    \left( 1 - \cos^2\alpha \right).
\end{equation}

\noindent The center of mass is lifted by this rotation. 
Therefore the gravitational potential energy per particle in 
the cluster increases by

\begin{equation}
\Delta E_{g} = N\ \frac{m g D}{2}\ \left( 1 - \cos\alpha \right).
\end{equation}

\noindent Setting $\alpha = 30$ degree and
using the definition of $f$ given in \Eq{f_def}, 
$\Delta E_{m} = \Delta E_{g}$ shows 
that clusters up to chain length

\begin{equation}
N_{max} \approx \left( 1 + \sqrt{3}/2 \right)\ f
\end{equation}

\noindent can peel off.

In order to check to what thickness $\nu_{max}$ such a cluster 
is stable with respect to splitting into subclusters with less chains, 
we compare the magnetic energy loss per unit chain length with
the gravitational energy gained, when a subcluster of $\nu$ chains
moves down by half a particle diameter relative to the rest of
the cluster. In this case we have

\begin{equation}
\Delta E_{m} = 
  \frac{\mu_{0}}{4\pi}\ \frac{S^2}{2 D^{3}},
\ \mathrm{and}\ 
\Delta E_{g} = \frac{m g D}{2}\ \nu.
\end{equation}

\noindent This shows that the dipole-dipole-interaction can only
prevent cluster splitting, if the cluster consists of less than

\begin{equation}
\nu_{max} \approx \frac{1}{6}\ f
\end{equation}

\noindent chains.

Based on the above results we can discuss the process of avalanche
formation. Already at small magnetizations the surface roughness 
allows for coherent rotation of larger
clusters up to chain length $N_{max}$.
For $f < f_c \approx 6$ clusters consisting 
of more than one chain of particles can easily dilate and will
disintegrate into isolated chains, forming a quasi-continuous flow.
For $f > f_c$ clusters consisting of up to $\nu_{max} > 1$ 
chains can fall. These results are close to the observations made on 
the simulations: The transition between the two regimes
was observed at $f_c \approx 7$.

The difference in avalanche formation in different regimes
can be clearly observed in avalanche duration 
to avalanche size relation. In all three simulation sets,
at given interparticle force ratios, for each avalanche
size we collected the measured avalanche durations and
calculated the corresponding average avalanche durations. 
We also examined the avalanche size and duration distributions.
Our results are summarized in the next two subsections.

\subsection{The granular regime} 

\begin{figure}[tbhp]
\begin{center}
\begin{tabular}{r}
\includegraphics{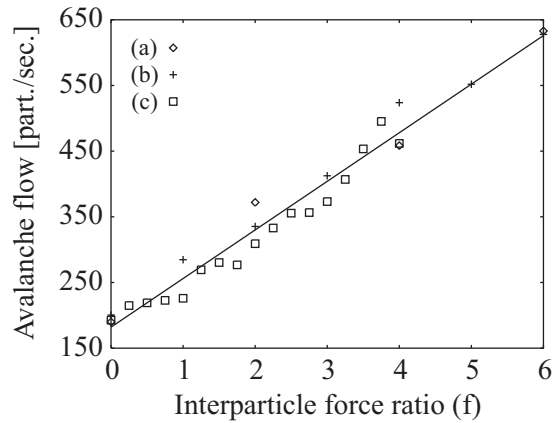} \\
\end{tabular}
\end{center}
\caption{
  Dependence of average avalanche flow on interparticle force ratio
  in granular regime.
  We examined three different simulation setups (see text for details).
  Firing the particles into the pile (i.e. dropping them from a given
  height) or placing them gently, and switching on or off the side
  wall effect gave no qualitative difference. 
  The avalanche flow is measured in particles per second.
}
\label{fig:duravrsmall}
\end{figure}

In the granular regime 
the avalanche sizes are proportional 
to the corresponding average avalanche durations. 
The proportionality factor (i.e. the ratio
of avalanche sizes and average avalanche durations) defines
an average avalanche flow, which increases linearly 
with $f$ (see \Fig{duravrsmall}). This linear dependence 
is explained by the fact that the avalanches
in granular regime are formed by quasi-continuous flows of
small particle chains, and the height of these chains
increases linearly with $f$.

Independent of the setup,
at zero magnetization the average avalanche flow is approximately
$182$ particles per second, and increases with approximately
$74$ particles per second per unit change of interparticle force 
ratio (see \Fig{duravrsmall}).

\begin{figure}[tbhp]
\begin{center}
\begin{tabular}{r}
\includegraphics{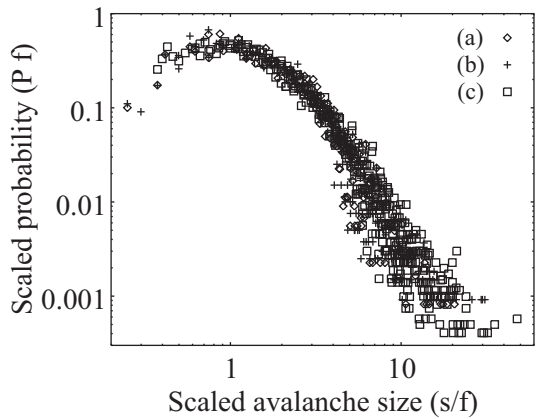} \\
\end{tabular}
\end{center}
\caption{
  Scaled avalanche size distribution in granular regime.
  A well-defined characteristic size can be observed.
  We examined three different simulation setups 
  (see text for details).
  The avalanche size distribution at interparticle
  force ratios $1 < f < 7$ are scaled together
  using the ansatz $P(s,f) = f^{-1} Q(s/f)$,
  where $s$ denotes avalanche sizes.
  From the scaling property, we conclude that, 
  the characteristic avalanche size
  increases linearly with $f$.
}
\label{fig:statshortsize}
\end{figure}

The avalanche size distribution at interparticle
force ratios $1 < f < 7$ can be scaled together
reasonably well (see \Fig{statshortsize})
using the ansatz

\begin{equation}
P(s,f) = f^{-1} Q(s/f),
\end{equation}

\noindent where $s$ denotes avalanche sizes, 
$P(s,f)$ is the probability associated with an avalanche size, 
and $Q(\cdot)$ is a function with integral $1$ on the 
$[0,+\infty)$ interval.

Based on the avalanche size distributions (see \Fig{statshortsize}), 
we argue that the magnetic cohesion introduces a well-defined
characteristic size in particle avalanches. 
From the scaling property, we conclude that, 
the characteristic avalanche size
increases linearly with the interparticle force ratio.
Qualitatively similar results were found in the experiments by 
Szalm\'as et al. \cite{szalmas-bme00}.

As both the characteristic avalanche size 
and the average avalanche flow increase linearly with $f$, 
the characteristic average avalanche duration 
(equal to the ratio of the former two)
in leading order is independent of $f$. The dependence
of the avalanche duration distribution on $f$ is contained
in higher order corrections which could not be captured by our
simulations. 

\subsection{The correlated regime} 

In correlated regime  
the avalanche durations are given 
by the free fall of long particle clusters, and
accordingly the square of the measured avalanche durations
are proportional to the length of the clusters,
and thus proportional to $f$.
At the same time, as it was already mentioned, the width of the
falling particle clusters is small, and thus the avalanche
sizes in leading order are proportional to the cluster length.
In consequence the avalanche
sizes are proportional to the square of the 
corresponding average avalanche durations, contrary to the linear
dependence found in granular regime.

As it was already mentioned,
the width of the falling particle clusters can take 
arbitrary values up to some well defined maximum.
In other words, the particle clusters can have different number of
\emph{layers} while having the same length. This introduces
large fluctuations in avalanche sizes. These fluctuations 
are proportional to the cluster length, and
thus proportional to $f$. 

\begin{figure}[tbhp]
\begin{center}
\begin{tabular}{r}
\includegraphics{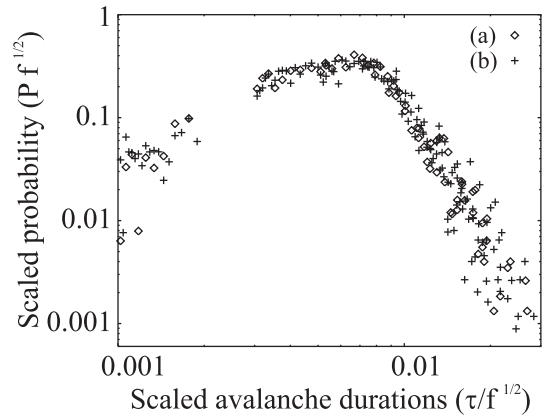} \\
\end{tabular}
\end{center}
\caption{
  Scaled avalanche duration distribution in correlated regime.
  A well-defined characteristic duration can be observed.
  We examined two simulation setups (see text for details).
  The avalanche duration distribution at interparticle
  force ratios $7 < f < 25$ are scaled together
  using the ansatz $P(\tau,f) = f^{-1/2} Q(\tau/f^{1/2})$,
  where $\tau$ denotes avalanche durations.
  From the scaling property, we conclude that, 
  the square of characteristic avalanche duration
  increases linearly with $f$.
}
\label{fig:statshorttime}
\end{figure}

The avalanche duration distributions at interparticle
force ratios $7 < f < 25$ can be scaled together 
reasonably well (see \Fig{statshorttime})
using the ansatz 

\begin{equation}
P(\tau,f) = f^{-1/2} Q(\tau/f^{1/2}),
\end{equation}

\noindent where $\tau$ denotes avalanche durations, 
$P(\tau,f)$ is the probability associated with an avalanche duration, 
and $Q(\cdot)$ is a function with integral $1$ on the 
$[0,+\infty)$ interval.

A well-defined characteristic duration can be observed
in \Fig{statshorttime}. Based on the scaling property, 
the square of the characteristic avalanche duration 
increases linearly with the interparticle force ratio.
As the square of the avalanche duration is proportional
to a corresponding cluster length, and a cluster length 
in leading order defines an avalanche size, 
we can conclude that
there is also a mean characteristic avalanche size,
which consequently is proportional to $f$. 
We could not find this avalanche size 
explicitly in the avalanche size distribution 
because of the large fluctuations 
of the same order of magnitude as the
mean value.

\section{Conclusions and discussion} 

We studied basic effects in two-dimensional granular piles
formed by magnetized particles in simulations similar to the experiments of 
Forsyth et al. \cite{forsyth-pre01,hutton-phd02}.
We measured the angle of repose, the surface roughness,
and the effect of magnetization on 
particle avalanching.
As a general result we can mention that
dropping the particles from a given
height or placing them gently, and switching on or off the 
effect of the side walls gave no qualitative difference. 

We found that both the angle of repose and the surface roughness 
exhibits linear dependence on the ratio $f$ of the maximum 
magnetic force at contact and the gravitational force.
As it was also mentioned by Forsyth et al. 
\cite{forsyth-pre01,hutton-phd02}, $f$
overestimates the effective cohesion because of the 
anisotropy of the magnetic interaction. According to this, 
the angle of repose increases much more slowly with $f$ than 
expected from stability criteria \cite{albert-pre97}
and experiments \cite{tegzes-pre99} on wet granular media.
The experimental results of Forsyth et al. and our
simulations are in good accordance,
though the angle of repose at zero magnetization in
our case was smaller.
The side wall model taking into consideration the effect of
the front and back walls of a Hele-Shaw cell arrangement, 
could increase the surface angle but did also 
introduce a stronger dependence of the angle of repose on $f$
than in the experiments of Forsyth et al. 
\cite{forsyth-pre01,hutton-phd02}.
Taking into consideration the
static and rolling friction of particles could probably reproduce more
closely the experimental results in this respect too.

Tegzes et al. \cite{tegzes-pre99} found in experimental studies
of wet granular media linear dependence of the angle 
of repose on the interparticle force ratio 
in the granular regime and an \emph{almost} linear dependence in
correlated regime with a slight curvature.
This bending could not be clearly
identified in our results and most probably would
require more accurate investigations at both small 
and large interparticle force ratios.

As reported by Tegzes et al. \cite{tegzes-pre03} 
in case of wet granular media there 
is a difference in avalanche formation at small 
and at large interparticle force ratios. We could also identify
a \emph{granular} and a \emph{correlated regime} in case of
magnetized particles. The granular regime is characterized
by quasi-continuous granular flows, 
while the correlated regime is characterized
by long particle clusters falling at the system's boundary.
The transition between the two regimes is not sharp.
In simulations we found that the transition 
is at $f_c \approx 7$, while calculations based on stability
criteria indicate a transition at $f_c \approx 6$ for the investigated
magnetic case.

In the granular regime the 
avalanche sizes are proportional to the corresponding average 
avalanche durations. According to this, there is a well defined
average granular flow characterizing the avalanches. We found that
this increases linearly with $f$. Analyzing the avalanche size
distributions, we also found that there is a
well-defined characteristic size in particle avalanches. 
Based on scaling properties of the avalanche size 
distributions, we argue that the characteristic avalanche size 
increases linearly with $f$. 
The characteristic average avalanche duration 
in leading order seems to be independent of $f$.
This dependence is contained
in higher order corrections which could not be captured by our
simulations. 

In the correlated regime the avalanche sizes are in leading 
order proportional to the square of the corresponding average 
avalanche durations. This is explained by the free fall of long
particle clusters.
The avalanche durations are defined by the length of the
falling particle clusters.
The width of the falling clusters is small, and 
can take arbitrary values up to a well defined maximum.
According to this the avalanche sizes are defined in leading order
by the cluster lengths, but there are large fluctuations which are 
also proportional to the cluster lengths. 
The cluster length increases linearly with $f$.
The avalanche duration distributions show
evidence of a characteristic avalanche duration,
indicating also a characteristic avalanche size.
We could not identify this avalanche size 
explicitly because of the large fluctuations 
in the avalanche size distribution.

Our results regarding the avalanche size distributions 
in both granular and correlated regime are very 
close to the experimental results of Tegzes et al. 
\cite{tegzes-pre03} on wet granular materials. 
They could identify characteristic
avalanche sizes in both regimes, and they state that the
avalanche size distributions in correlated regime are
\emph{broad} and both small and large avalanches may occur.
This seems to confirm our finding that in correlated regime
there are large avalanche sizes fluctuations.

The results on avalanche sizes and durations may slightly depend on 
the chosen time scale on which the avalanches are observed, however 
we argue that much more coarser or finer time scale will both lead to 
non-physical results, while small correction in the time scale
will not lead to qualitative difference.

\section{Acknowledgments}


This research was carried out within the framework of the
``Center for Applied Mathematics and Computational Physics'' of the
BUTE, and it was supported by BMBF, grant HUN 02/011, and Hungarian 
Grant OTKA T035028.

\bigskip

\bibliography{magpart}

\end{document}